# Network connectivity dynamics affect the evolution of culturally transmitted variants

**José Segovia Martín**

Cognitive Science and Language (CCiL)

Universität Autónoma de Barcelona

jose.segovia@e-campus.uab.cat

**Bradley Walker**

School of Psychological Science

University of Western Australia

bradley.walker@uwa.edu.au

**Nicolas Fay**

School of Psychological Science

University of Western Australia

nicolas.fay@uwa.edu.au

**Mónica Tamariz**

Department of Psychology

Heriot Watt University

m.tamariz@hw.ac.uk

## Abstract

The distribution of cultural variants in a population is shaped by both neutral evolutionary dynamics and by selection pressures, which include several individual cognitive biases, demographic factors and social network structures. The temporal dynamics of social network connectivity, i.e. the order in which individuals in a population interact with each other, has been largely unexplored. In this paper we investigate how, in a fully connected social network, connectivity dynamics, alone and in interaction with different cognitive biases, affect the evolution of cultural variants. Using agent-based computer simulations, we manipulate population connectivity dynamics (early, middle and late full-population connectivity); content bias, or a preference for high-quality variants; coordination bias, or whether agents tend to use self-produced variants (egocentric bias), or to switch to variants observed in others (allocentric bias); and memory size, or the number of items that agents can store in their memory. We show that connectivity dynamics affect the time-course of variant spread, with lower connectivity slowing down convergence of the population onto a single cultural variant. We also show that, compared to a neutral evolutionary model, content bias accelerates convergence and amplifies the effects of connectivity dynamics, whilst larger memory size and coordination bias, especially egocentric bias, slow down convergence. Furthermore, connectivity dynamics affect the frequency of high quality variants (adaptiveness), with late connectivity populations showing bursts of rapid change in adaptiveness followed by periods of relatively slower change, and early connectivity populations following a single-peak evolutionary dynamic. In this way, we provide for the first time a direct connection between the order of agents' interactions and punctuational evolution. These results confirm, complement and extend formal and experimental findings suggesting that connectivity dynamics critically affect the way in which cultural evolution proceeds.[1]

---

[1] Code to run the analyses described in this paper is available at:
https://github.com/jsegoviamartin/network_connectivity_dynamics_model

# 1. Introduction

Human life is shaped by our culture, that is, by socially transmitted information that determines our behaviour, beliefs, attitudes and values (Boyd & Richerson, 2005). Cultural variants such as technology, language and beliefs propagate in populations following evolutionary dynamics (Cavalli-Sforza & Feldman 1981; Boyd & Richerson, 1988; Neiman 1995) -- individuals inherit cultural traits from ancestors or peers and occasionally generate new trait variants. Over generations, cultures evolve: some variants are lost, while others spread in a population, sometimes to the point of fixation, when we can say the population has *converged* on a variant.

Drift models have been used to explain cultural evolution (Bentley, Hahn, & Shennan, 2004), and the evolution of human communication systems (e.g. Reali, & Griffiths, 2010; Blythe, 2011). When evolution is neutral (***drift*** dynamics), all variants are equally likely to be adopted by individuals, and therefore to propagate in the population. In the absence of innovation, drift leads, over generations, to convergence on a single variant. Drift may explain the propagation of cultural variants including baby names, pottery decorations and patents (Bentley, Hahn, & Shennan, 2004), dog breeds (Herzog, Bentley, & Hahn, 2004) and some diachronic changes in language (Kroch, 1989; De Graff, 2001; Komarova, & Nowak, 2003). Drift dynamics are dependent on demographic factors: for example, the influence of stochastic variation is higher for small populations, as has been shown in population genetics (Frankham, Briscoe & Ballou, 2002). Similarly, small populations may converge faster than large ones in some cultural variants, leading to faster extinction or selection of technological skills (Henrich, 2004). Drift models can be used as null models against which other models can be tested (Reali, & Griffiths, 2010; Neiman 1995; Lipo, Madsen, Dunnell & Hunt, 1997; Shennan & Wilkinson 2001; Hahn & Bentley 2003).

Some variants spread more rapidly than others. In these cases, evolution is not neutral, but subject to biases, or **selection pressures**. *Content bias,* also termed direct bias by Boyd & Richerson (1985), relates to intrinsic properties of traits, and results in the more learnable, efficient or effective variants having a higher probability of being adopted by others (Henrich, & McElreath, 2007; Vale, Flynn, Kendal, Rawlings, Hopper, Schapiro, Lambeth & Kendal.2017; Hagen & Hammerstein, 2006), and therefore spreading faster through a population than a neutral, drift model would predict (Gong et al. 2012; Tamariz, Ellison, Barr, & Fay, 2014). *Coordination biases* may involve a preference to use variants we have used before (egocentric bias) or variants produced by others (allocentric bias). In communicative tasks, for instance, an allocentric bias is observed, since speakers tend to adopt labels used by their interlocutors (Pickering & Garrod, 2004; Garrod & Pickering, 2007), which in turn favours cooperation and coordination (Fusaroli, Bahrami, Olsen, Roepstorff, Rees, Frith & Tylén, 2012; Fusaroli & Tylén, 2016). Content and coordination biases also interact with each other: egocentric bias maintains variation, which improves the chances that content bias will select for the most adaptive variant in a population (Walker, Segovia-Martin, Tamariz & Fay, under review). Cultural transmission is also affected by the memory record of cultural variants. Some authors claim that the type of variation that learners produce can be explained by memory limitations: for example, memory can affect language regularization (Hudson Kam & Newport, 2005, 2009; Hudson Kam & Chang, 2009), compressibility (Chater & Vitanyi, 2003) or conventionalization (Tamariz & Kirby, 2015). In general, memory limitations reduce variation (Ferdinand, Thompson, Kirby & Smith, 2013; Tamariz & Kirby, 2015).

The effects of content and coordination biases on variant propagation were studied by Tamariz et al. (2014), who constructed a parameterized model of cultural variant transmission to analyze the patterns of variant spread obtained in an experimental study by Fay et al. (2010). Fay et al. (2010) had groups of eight individuals play a Pictionary-like communication game in pairs. During each game, the 'director' produced a drawing to represent each of 16 meanings, one at a time. For each of these, the 'matcher' tried to guess which of 20 possible meanings the director was trying to communicate. Participants played this game six times with each partner, with roles reversing for each game, so each participant drew and matched each meaning three times with each partner. For a given meaning, directors could invent their own ways to depict the meaning (that is, produce a novel variant) or produce a variant that was produced by a partner or by themselves in a previous game. After six games, participants swapped partners within their group and played another six games with their new partner. This partner-swapping was repeated until the populations were fully connected, with every participant having played with every other participant. This meant that, by the end of the experimental simulation, for each meaning, a particular variant could spread to the whole population. In other words, the population could converge on a single variant for each meaning. Tamariz et al. (2014) found that this convergence was best explained by a combination of egocentric bias and content bias, where participants would stick with variants they had produced previously, unless they encountered a better variant, in which case they would switch to that.

Demographic factors also add a selection pressure (Richerson & Boyd, 2005; Mesoudi, Whiten, & Laland, 2006; Henrich, 2004; Mesoudi, 2011; Shennan, 2001; Vaesen, 2012). The degree of adaptiveness, complexity and cumulative cultural evolution of cultural variants positively correlates with **population size** (Shennan, 2001; Kline & Boyd, 2010; Henrich, 2004; Kobayashi & Aoki, 2012; Derex, Beugin, Godelle & Raymond, 2013; Kempe & Mesoudi, 2016), with the degree of **contact** and migration between populations (Powell, Shennan & Thomas, 2009; Muthukrishna, Shulman, Vasilescu & Henrich, 2014; Creanza, Kolodny & Feldman, 2017) and with the **structure of the social network** (Olfati-Saber & Murray, 2004; Mueller-Frank, 2013; Lee, Stabler, & Taylor, 2005; Lupyan & Dale, 2010; Gong, Minett & Wang, 2008). An additional demographic variable, namely the connectivity between individuals within or across populations, also enhances adaptiveness and complexity because it affects the degree of diversity each individual has access to (Shennan, 2001; Henrich, 2004; Powell, Shennan & Thomas 2009; Kobayashi & Aoki, 2012). However, a recent study suggests that access to diversity is not the only variable at play, and that if we take into consideration the potential for an innovation to be adopted and spread, then an intermediate degree of connectivity may be optimal for cumulative culture, as too much connectivity stifles innovation, whilst too little cannot maintain complex traits (Derex & Boyd 2016; Derex, Perreault & Boyd, 2018).

In the evolution and history of human populations, the structure of the population might have played an important role in cultural change in ancestral and historical periods (Derex & Boyd, 2016). Critically, it should be noted that population fragmentation and cultural isolation have been identified as crucial factors to explain the spread of cultural variants such as high quality ideas (Björk & Magnusson, 2009), technology (Hovers & Belfer-Cohen, 2006) or research (March, 2005). Furthermore, it has been suggested that *inter-population connectivity* may be more than just a simple reflection of cultural accumulation, and that it may be a critical driver of cultural change (Creanza, Kolodny, & Feldman, 2017).

An as-yet unexplored aspect of social network research is the effect of the order in which connections between individuals unfold over time. The most basic network topology is a fully connected network, in which all nodes are interconnected. In a population of individuals, this means that, over time, each individual interacts with every other individual. However, the same fully connected network may follow different temporal patterns of connectivity, and this may have consequences for variant spread. For instance, a pattern may lead to the temporary isolation of one or more sub-populations. In such small sub-populations, drift can reduce diversity, disproportionately favouring variants that happen to be present in the population, and which are not necessarily adaptive (Henrich 2004). A different connectivity pattern may never yield pockets of isolation, which would lead to different evolutionary dynamics. In this paper, we address how different temporal patterns of connectivity in a fully connected social network (which we call 'connectivity dynamics'), alone and combined with content and coordination biases, affect the spread of cultural variants in a population.

Tamariz et al.'s (2014) study was designed to test whether the observed variant distributions obtained by Fay et al. (2010) were best explained by neutral drift or showed evidence of selection and adaptation mediated by content and/or coordination biases. Their results indicated an interaction of both biases: participants displayed egocentric bias and content bias; they tended to produce the variants they had previously produced themselves except when they encountered a better (simpler, cleverer, etc.) variant (through mutation/innovation or via a partner), in which case they tended to adopt it. Additionally, participants seemed to have 'short memory' and tended to produce mostly variants that they had seen or produced in the preceding couple of games. For the present study, we extend Tamariz et al.'s (2014) study in two important ways: first, as well as levels of content and coordination biases and memory size, we manipulate the population connectivity dynamics. Second, our model is productive -- it is not designed to fit existing variant distributions from experiments and datasets, but to produce controlled variant distributions. These two innovations allow us to establish causal links between properties of the individuals (content and coordination biases, memory size) and of the population (connectivity dynamics and population size) on one hand, and properties of the culture (evolution and adaptiveness of variants) on the other.

It is important to note that even though previous work in the field has frequently used fully connected networks (Komarova, Niyogi, & Nowak, 2001; Fay, Garrod, Roberts & Swoboda, 2010; Tamariz et al., 2014), this type of network topology is unrealistic because it restricts the interaction between agents to a particular pattern of interconnectedness, reducing the complexity of the system. Therefore, population connectivity dynamics might play a different role in other networks, such as scale-free networks, which are more common in nature and in human cultural systems (Barabási, Albert & Jeong, 1999; Barabâsi, Jeong, Néda, Ravasz, Schubert & Vicsek, 2002). Furthermore, it is well established that people do not contribute equally to group discussions, leading to different degrees of network connectivity (Stasser & Taylor, 1991; Fay, Garrod & Carletta, 2000). Both factors might potentially motivate future extensions of our current investigation.

We are also aware of a number of studies that in the fields of computer science and algebraic graph theory have addressed consensus problems under a variety of assumptions about the network topology by providing convergence analysis (Olfati-Saber & Murray, 2004, Mueller-Frank, 2013). Our paper addresses convergence problems from a cultural evolutionary perspective by simulating how the interplay between individual biases and the network dynamic evolves over time. By doing this, we provide a direct connection between convergence and the network dynamic for a fixed range of combinations of individual cognitive biases. This computational modeling approach is especially useful as an exploratory method in the field of cultural evolution, since it allows us to establish links between population connectivity dynamics and evolutionary trajectories in a straightforward way.

Two important questions are addressed in this paper. First, to what extent do coordination bias and content bias affect the evolution of cultural variants? These analyses replicate and extend previous work. Second, and our primary focus, do connectivity dynamics affect the rate of convergence of variants in a population, and do they interact with the cognitive biases and memory size, modifying the rate of convergence and the adaptiveness of cultural variants during cultural evolution?

## 2. Methods

The research questions above are addressed by simulating the evolution of a pool of variants under controlled conditions. We manipulated the values of connectivity dynamics and cognitive biases, and quantify the resulting changes in the evolution of the cultural variants.

### 2.1 The agent-based model

We constructed an agent-based computer model inspired by Tamariz et al. (2014), in which simulated agents played recurring games in pairs. We simulated pairwise interactive micro-societies of 8, 16 and 32 agents, allowing us to track all agent pairings. In each round, each player produced a variant from memory, and both agents stored the two variants in memory. They labeled the variants as produced by self or produced by partner. (At round 0, each agent produced a unique signal variant). At each round, agents switched partners so that by the end of the final round every agent had played with every other agent. In successive rounds, each agent produced either a variant from its memory record, or a novel one, in which case we speak of mutation. Each variant in the memory record was annotated as either produced by the agent ($E$ for *ego*) or by one of their partners ($A$ for *allo*). The relative frequencies of variants in egocentric memory (or agent's history) $h_{|E,m}$ defined the egocentric initial distribution $f(h_{|E,m})$, and in allocentric memory $h_{|A,m}$ the allocentric distribution $f(h_{|A,m})$. Thus, $f$ mapped each item in a list with its relative frequency. In order to choose which variant to produce at each round, agents sampled from the probability distribution yielded by a model defined by the following parameters:

a) Population connectivity dynamics. This reflects the order in which the agents were paired with each other. Different orderings yielded different levels of sub-population isolation at different times (Fig. 1). For example, in our 8-agent micro-societies, three different connectivity dynamics can be described with reference to how many agents could potentially share the same variant by round 3: in the early connectivity condition, all 8 agents could share the same variant by round 3; in the mid connectivity condition, 6 agents could share same variant by round 3; in the late connectivity condition, 4 agents could share the same variant by round 3. We examined 3 connectivity dynamics (early, mid, late) in 8-agent micro-societies, and 2 connectivity dynamics (early, late) in 16-and 32-agent micro-societies. We did not include mid connectivity in the larger micro-societies due to the wide range of possible permutations.

A

| Round | Early | | | | Round | Mid | | | | Round | Late | | | |
|---|---|---|---|---|---|---|---|---|---|---|---|---|---|---|
| 1 | 1&2 | 3&4 | 5&6 | 7&8 | 1 | 1&2 | 3&4 | 5&6 | 7&8 | 1 | 1&2 | 3&4 | 5&6 | 7&8 |
| 2 | 1&4 | 3&2 | 5&8 | 7&6 | 2 | 1&4 | 2&7 | 3&6 | 5&8 | 2 | 1&4 | 3&2 | 5&8 | 6&7 |
| 3 | 1&6 | 3&8 | 5&2 | 7&4 | 3 | 1&6 | 4&7 | 2&5 | 3&8 | 3 | 1&3 | 2&4 | 5&7 | 6&8 |
| 4 | 1&8 | 3&6 | 5&4 | 7&2 | 4 | 1&5 | 3&7 | 2&6 | 4&8 | 4 | 1&5 | 2&6 | 3&7 | 4&8 |
| 5 | 1&3 | 2&4 | 5&7 | 6&8 | 5 | 1&7 | 5&3 | 2&8 | 6&4 | 5 | 1&6 | 3&8 | 5&2 | 7&4 |
| 6 | 1&5 | 2&6 | 3&7 | 4&8 | 6 | 1&8 | 3&2 | 7&6 | 5&4 | 6 | 1&7 | 2&8 | 3&5 | 4&6 |
| 7 | 1&7 | 2&8 | 3&5 | 4&6 | 7 | 1&3 | 5&7 | 2&4 | 6&8 | 7 | 1&8 | 3&6 | 5&4 | 7&2 |

B

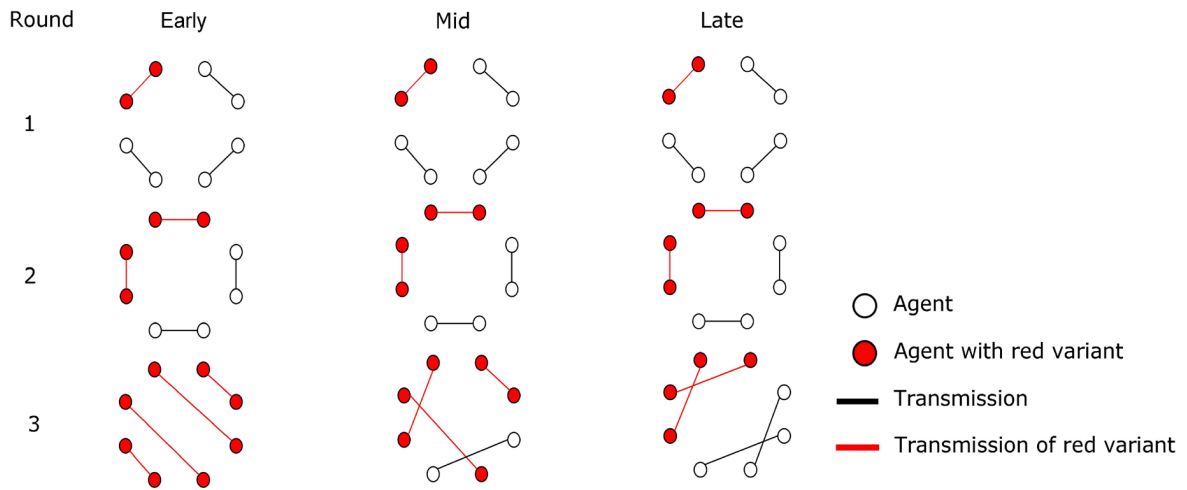

Figure 1. A. Examples of pair shuffling for each type of connectivity dynamic, in a micro-society of 8 agents. In red, potential spread of a variant produced by agent 1 in round 1. B. Three different network connectivity dynamics can be described within a pairwise interaction account for a dynamic fully connected network of 8 agents. By round 3, potentially 8 (in the early connectivity dynamic), 6 (in the mid connectivity dynamic) or 4 (in the late connectivity dynamic) agents share the red variant.

<u>b) Coordination bias</u> (*c*) fixed the likelihood of a variant being produced depending on whether it originated in the other partners or the agent itself. It took values from 0 (fully egocentric: preferring self-produced variants over other-produced variants) to 1 (fully allocentric: preferring other-produced variants over self-produced variants). When coordination bias was 0.5, we had a drift model: variants in $h_{|E,m}$ and in $h_{|A,m}$ were equally likely to be produced. Coordination bias values from 0 to 1 in steps of 0.1 were examined.

<u>c) Variant quality</u> ($s$) corresponded to the intrinsic value of each communicative variant and it indicated to what extent the variant was 'adaptive', or preferred over the other variants ($s$ was 1 if the signal was preferred over the others, 0 otherwise). (For adaptiveness measures, the frequency of high quality variants ($s=1$) were considered., see below.)

<u>d) Content bias</u> ($\beta$) encompassed two parameters ($b$, $d$). Parameter $b$ was the agents' sensitivity to variant quality ($s$), and ranged from 0 -not sensitive at all- to 1 -fully sensitive- in steps of 0.1. Parameter $d$ specified whether the variant was in the agent's memory record ($d$ was 1 if the variant was in memory, 0 otherwise). Parameter $\beta$ was equal to $b \times d$. Thus, content bias ($\beta$) assigned a value from 0 to 1 to each variant. When content bias was 0, we had a drift model. Content bias values from 0 to 1 in steps of 0.1 were examined.

<u>e) Memory size.</u> We manipulated memory size ($m$) by limiting agents' access to their memory store. At each round, when an agent had to produce a variant, it could only choose between the variants that were stored in the preceding 1, 3, 5 or 7 rounds.

<u>f) Mutation rate</u> ($\mu$). Agents could generate novel variants. We fixed the mutation probability at 0.02 by using a flat distribution $\varphi(x)$ weighted by mutation rate, which meant that 98% of variant choices would reflect the combined distribution (probability distribution yielded by all the parameters described above), while 2% would be a random choice among all 8 initial variants. The probability level selected reflects the mutation rate found in Fay et al.'s (2010) experimental data (Tamariz et al. 2014).

For each round in the simulation, the model yielded a probability distribution of variants ($x$) for a given history ($h$) of previous rounds, according to the following equation. (The overbar denotes the probabilistic complement.)

$$P(x|h) = \bar{\mu}\bar{\beta}\bar{c}f(x|h_{|E,m}) + \bar{\mu}\bar{\beta}cf(x|h_{|A,m}) + \bar{\mu}\beta s + \mu\varphi(x)$$

(Equation 1)

The data structure included 1452 different parameter value combinations in each round. For each parameter combination, we ran the simulation 1000 times. The results below show the average and standard deviations of the number of runs of each parameter combination examined. A table with all parameters and levels examined can be found in the supplementary material (Table 1S).

## 2.2 Quantifying convergence and adaptiveness

At its most general level, evolution is defined as a change in the frequencies of different variants in a population over time. We are therefore interested in the composition of the pool of 8, 16 or 32 variants produced by the agents at each round, and how it changes over rounds. We examine the level of <u>convergence</u> in this pool, or the extent to which agents used the same variant. Following others (e.g. Feher, Wonnacott & Smith, 2016; Smith & Wonnacott, 2010), we quantify convergence using the information-theoretic notion of entropy *(H)* (Shannon, 1948):

$$H(V) = -\sum_{v_i \in V} p(v_i) log_2 p(v_i)$$

(Equation 2)

where *V* corresponds to the set of variants, and $p(v_i)$ is the probability of $i^{th}$ variant in that set. High entropy corresponds to low convergence.

Evolution, even by drift, may increase convergence (and decrease entropy), as random sampling at each round gradually eliminates variants from the pool (and our low level of mutation is not enough to compensate for that). For example, at round 0, where each agent produced its own unique variant, the probability distribution over the 8 variants was flat (each variant had a probability of ⅛) and the entropy was maximal (*H* = 3 bits). Over time, as agents converged, entropy would decrease; if, say, by round 7, the probability of 1st variant was 0.75, the probability of 2nd variant was 0.25 and the probability of the remaining variants was 0, the entropy would be 0.811 bits.

In our model, therefore, convergence is the change (specifically, the decrease) in entropy over rounds. To better understand the mechanics of evolutionary algorithms, a number of studies have investigated <u>time to convergence</u> (*TC*). The number of rounds until convergence has been used to analyze convergence properties of genetic algorithms in studies about population sizing, network structures and theory of convergence (e.g. Mueller-Frank, 2013; Olfati-Saber & Murray, 2004; Pelikan, M., Goldberg, D. E., & Cantú-Paz, E., 2000). This additional measure is important because it allows us to predict more accurately the moment at which one population will reach convergence under different conditions. Additionally, it gives us more information about how relevant agents' choices were in the firsts rounds. Therefore, in some analyses we will also present time to convergence (TC) or the number of rounds it takes for the population to reach full convergence (defined as H = 0 bit) for the first time.

We calculated the <u>adaptiveness</u> *(A)* of the cultural system at each round (*t*) as the frequency of high-quality variants (quality is measured by *s*, see above) at that round,

$$A(t) = \frac{n(t)}{N(t)}$$

(Equation 3)

where *n(t)* is the number of high quality variants at round *t* and *N(t)* is the total set of variants produced in round *t*.

A number of researchers have used the notion of relative fitness and change in genotype frequencies in the Wright-Fisher model (Fisher, 1930; Wright, 1931) as an equivalent to Bayesian models, establishing a connection between biological and cultural evolution (Reali, & Griffiths, 2010). Other researchers on cultural evolution have developed models that link demography and cultural adaptiveness, using a variety of mathematical approaches (Mesoudi, 2011, Shennan, 2001). These models describe how a trait changes in frequency over time. For example, the average change in the frequency of cultural variants $\Delta \bar{z}$ is given by the Price equation as described in Henrich (2004):

$$\Delta \bar{z} = Cov(f, z) + E(f \Delta z)$$

(Equation 4)

where the first term of Equation (4) represents change due to selection (selective transmission) and the second term represents noisy interference. In order to make these equations helpful for studying cultural evolution, researchers usually make a series of tractability assumptions depending on the specific conditions of their study, e.g. all agents copy the most skilled agent in Henrich (2004). Following Vaesen (2012) Equation (4) can be reduced to:

$$\Delta \bar{z} = z_h - \bar{z} + \Delta \bar{z}_h$$

where $z_h$ is the frequency of a variant in the subsequent generation or $z(t + 1)$, and $\bar{z}$ corresponds to the z value of the earlier generation or $z(t)$.

Two considerations apply to our study when we calculate average change in high quality variants. First, we use relative fitness equations to account for the adaptiveness of high quality variants, that is to say, we only consider the frequency of cultural variants having $s = 1$. Second, we assume that cultural variants are distributed, and therefore transmitted at each round, according to our parametrized model, as defined in Equation (1) (see Methods 2.1). Thus, in our case, the change in the adaptiveness ($\Delta A$) of high quality variants **due to selection** follows immediately from our definition of adaptiveness. This tractability assumption simplifies our equation considerably, because now the change in adaptiveness equation reduces to:

$$\Delta A = A(t+1) - A(t)$$

(Equation 5)

where change in adaptiveness $\Delta A$ takes the difference between the adaptiveness in the subsequent round $A(t+1)$ and the adaptiveness in the earlier round $A(t)$. Therefore, a change in adaptiveness above 0 ($\Delta A > 0$) indicates that the fitness of high quality variants produced by agents increased from one round to the next. When $\Delta A = 0$, variant frequency was stable from round to round.

## 3. Results

In the next section we show a summary of the results of two selection models (content bias and coordination bias) against a drift model, and how they interact with each other. Next, we show the effects of memory limitations. These first two sections replicate and extend previous work (Tamariz et al. 2014; Walker et al. *under review*). Figures for these sections and additional analyses on conditional entropy distributions can be found in the supplementary materials. In the last section, we focus on the main original contribution of this paper: the effects of population connectivity dynamics on entropy, time to convergence and change in adaptiveness of the cultural system from round to round. We also pay special attention to the interplay between connectivity and two strong drivers of convergence: content bias and population size. We use linear and non-linear regressions to fit models to our data to establish the relationships between variables (see supplementary material). However, following White et al. (2014), we use frequentist statistical models only to calculate effect sizes in our multifactorial simulations, but we do not report *p*-values, which can be meaningless when applied to simulation model output. In the following analyses we show mean values and standard deviations (Mean ± SD).

### 3.1 Cognitive biases

We ran simulations manipulating the level of content bias. As seen in Figures 1S and 2S (see supplementary materials), when compared with a drift model (content bias = 0, red bars in Fig. 1S), content bias increased convergence (decreasing entropy). Mean entropy was greatest when content bias was 0 (2.451 ± 0.448 bits) and lowest when content bias was 1 (1.020 ± 1.136 bits). Similarly, when keeping all other parameters at drift level (coordination bias = 0.5), mean entropy was greatest when content bias was 0 (2.277 ± 0.478 bits) and lowest when content bias was 1 (1.015 ± 1.137 bits).

Turning now to coordination bias, egocentric bias, slowed entropy decline (Fig. 3S, 4S). Mean entropy was greatest when coordination bias was 0 (strongest egocentric bias) (2.013 ± 1.060 bits) and lowest when coordination bias was 0.5 (neutral coordination) (1.768 ± 0.956 bits). Similarly, when keeping a neutral content bias (content bias = 0), mean entropy was greatest when coordination bias was fully egocentric, 0 (2.906 ± 0.171 bits) and lowest when coordination bias was neutral, 0.5 (2.277 ± 0.478 bits).

Both content bias and coordination bias had effects on convergence that differ from a drift model (Fig 5S). The effect of coordination bias on entropy was different for each level of content bias, revealing an interaction. Average entropy was greatest when content bias was 0 and coordination bias was also 0 (weakest content bias and strongest egocentric bias) (2.906 ± 0.171 bits) and lowest when content bias was 1 and coordination bias was 0.5 (1.015 ± 1.136 bits). When agent behavior was strongly content-biased, the rate of convergence increased, masking the effect of the coordination bias. Conversely, weaker content biases allowed coordination bias to show its effect on convergence, which can be characterized by a distinctive asymmetric distribution. The slowing effect of the coordination bias on the rate of convergence becomes hidden as content bias rises (Fig. 6S, 7S).

Regarding adaptiveness, when compared with a drift model, this measure only increased in content-biased models (Fig. 5SA). In other words, a certain level of content bias, with or without coordination bias, was required to increase adaptiveness.

## 3.2 Memory

Memory size increased entropy and therefore decreased convergence. Based on the global dataset, the average entropy was greatest in the absence of memory limitations, when agents kept in memory all the variants they had been exposed to (1.935 ± 0.920 bits). In contrast, when we limited agents' memory to the most recent 5 rounds (1.920 ± 0.930 bits), 3 rounds (1.834 ± 0.990 bits) or 1 round (1.675 ± 1.081 bits), entropy decreased noticeably.

This effect of memory on entropy was amplified by content bias. However, this was more noticeable for intermediate values of content bias (content bias = 0.5). When content bias was strongest, memory effects tended to be masked by a floor effect at the lower end of the entropy distribution in the last rounds. On the other hand, when content bias was neutral (content bias = 0), memory effects were greater for neutral values of coordination bias (of around 0.5), when compared with strongly egocentric or allocentric bias (Fig. 8S).

Similarly, adaptiveness of the cultural system increased with memory limitation. When compared with a complete memory model, memory limitations always increased adaptiveness, reaching highest values when memory was limited to 1 round (Fig. 8SA).

### 3.3 Connectivity dynamic of the population

The connectivity dynamic affected entropy. Convergence was delayed in populations that took longer to reach full connectivity. But this socio-structural effect only manifested under certain conditions related to the cognition of individual agents. When running simulations using a drift model, entropy remained similar for all levels of connectivity (Fig. 2). However, increasing content bias in the agents revealed a substantial effect of the connectivity dynamic on convergence. Mean entropy differences between conditions were greatest at generation 4: $0.805 \pm 0.730$ bits under late connectivity, $0.464 \pm 0.498$ bits under mid connectivity, $0.133 \pm 0.286$ bits under early connectivity. This effect was somewhat masked by a floor effect in later generations when content bias was stronger (entropy scores tended to cluster at the lower end of the distribution because it was not possible to have entropy values lower than 0) (Fig. 2). That is why mean entropy differences between conditions remain more noticeable over rounds as long as entropy does not drop to values close to 0, that is, for high-intermediate levels of content bias.

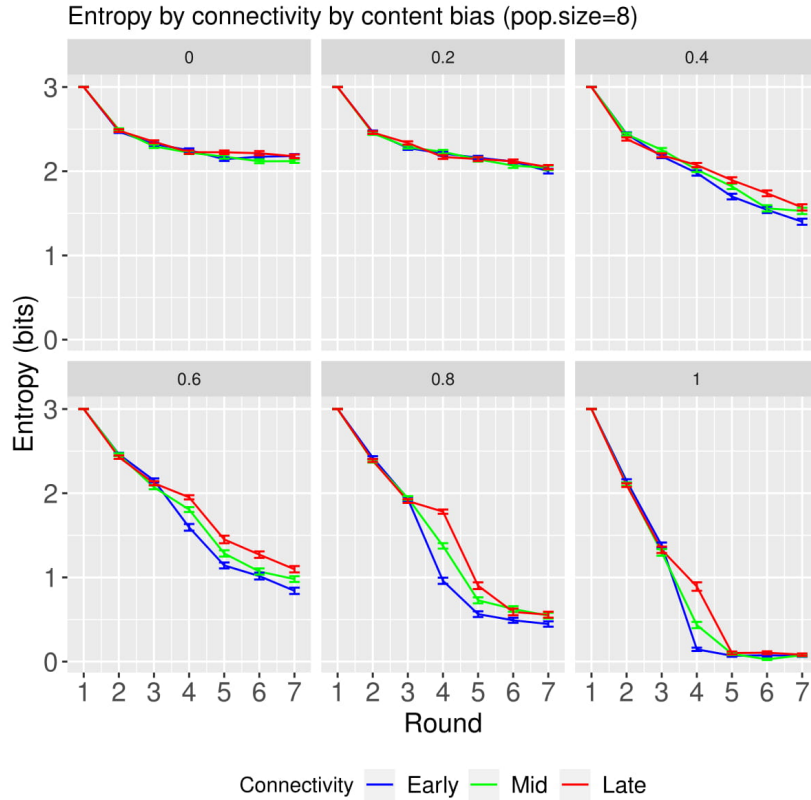

Figure 2. Entropy (*H*) averaged over each level of connectivity (Early, Mid, Late) and content bias (here we show results for 0, 0.2, 0.4, 0.6, 0.8 and 1) when coordination = 0.5. The *x*-axis represents rounds from 1 to 7, and the *y*-axis represents mean entropy in bits (normalized values to the range of min. 0, max. 1). Drift (top-left). We ran 1000 simulations for each parameter combination using 8-agent micro-societies. Error bars indicate 95% CIs. A plot with all levels of content bias can be found in supplementary materials (Fig. 9S).

The effect imposed by the connectivity dynamic was more pronounced in larger populations, where convergence was slower – we can observe the delay enforced by the late connectivity dynamic for a larger number of rounds when we increase population size (Fig. 3). In content-biased populations, mean entropy differences between conditions remained significantly high in the long term until late connectivity populations reached their equilibrium several rounds after early connectivity populations reached theirs (e.g. when content bias is 0.8, the relative difference between conditions remained above 10% for more than 6 rounds in 16-agent micro-societies, and for more than 11 rounds in 32-agent micro-societies). Furthermore, in intermediate content-biased populations, these differences tended to be irreversibly fixed (Fig. 3). Thus, both content bias, by strengthening the selection of high quality variants, and population size, by lengthening the time to convergence, amplified the effect of the connectivity dynamic, and this in turn resulted in a deep alteration of the convergence's evolutionary trajectory. In these scenarios, late connectivity populations clearly show periods of rapid convergence followed by periods of relatively slower convergence, resembling punctuational evolutionary dynamics. In contrast, convergence in early connectivity populations was not affected by these evolutionary bursts and tended to be shaped by a monotonic sigmoid curve.

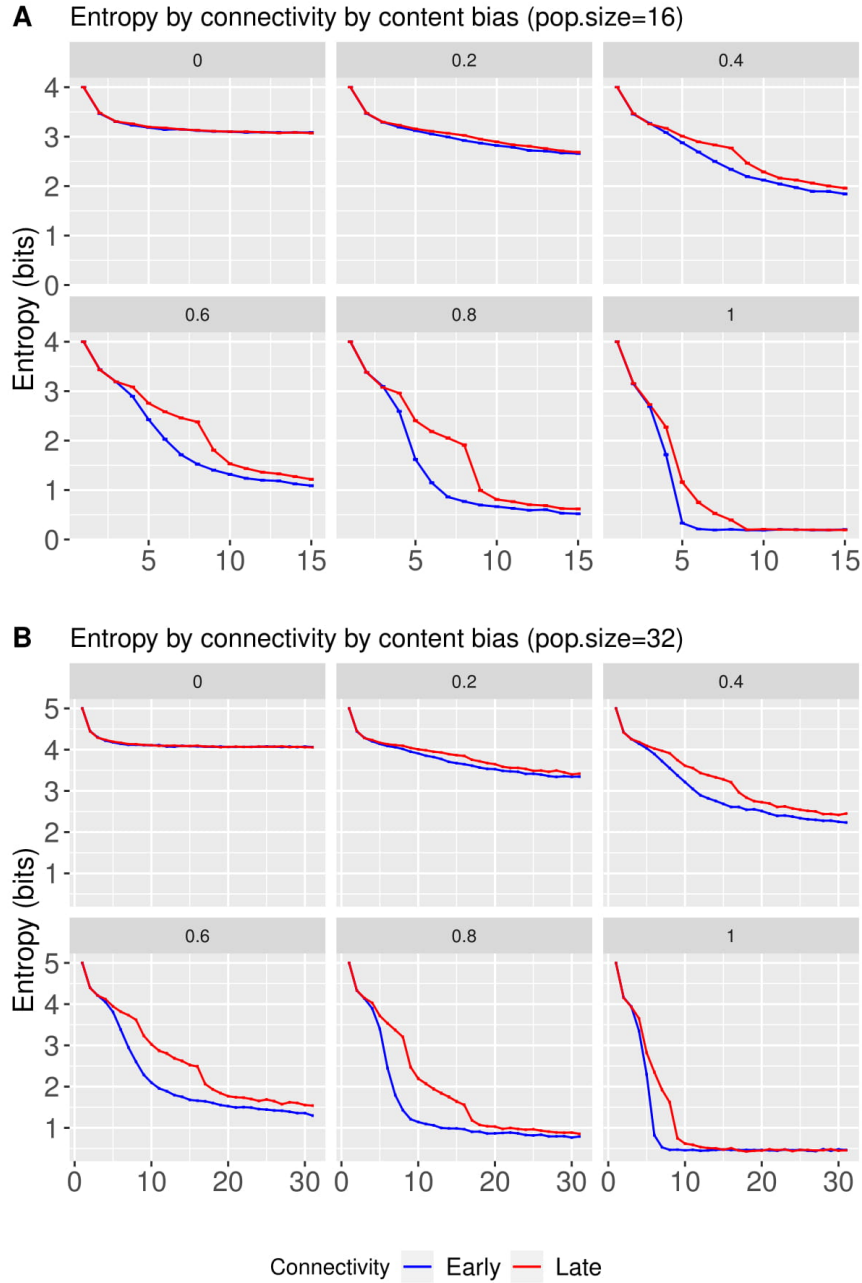

Figure 3. Entropy (H) averaged over each level of connectivity and content bias, for population size=16 (A) and population size=32 (B). The *x*-axis represents rounds, and the *y*-axis represents entropy in bits. Drift models are shown

in the top-left (content bias = 0). We ran 1000 simulations for each parameter combination. Error bars indicate 95% CIs. Plots with all levels of content bias can be found in supplementary materials (Fig. 10S, 11S, 12S, 13S).

We also examined the effect of connectivity on the **time to convergence**. The time required for a population to reach full convergence ($H = 0$) was longer in the late connectivity condition when compared to the mid and early connectivity conditions (Fig. 14S). Furthermore, content bias accelerated convergence in all conditions when compared with a drift model (content bias=0 and coordination bias=0.5) and amplified the effect of the connectivity dynamic (Fig. 14S). Coordination bias and memory do not seem to interact with the connectivity dynamic when it comes to explaining time to convergence (supplementary materials Fig. 15S, 16S).

Similarly, the adaptiveness *(A)* of the cultural system increased more rapidly in populations with early connectivity. The change in adaptiveness of high quality variants remained above 0 across rounds, indicating that the proportion of high quality variants always increased from round to round. However, changes in adaptiveness followed different patterns in populations with early, mid and late connectivity. Populations with late connectivity evolved in punctuated bursts of change followed by periods of slower change. For instance, in 8-agent micro-societies, at least 2 rapid bursts of change in the proportion of high quality variants can be observed before the population became a fully connected network in 7 rounds (Fig. 4). As above, these patterns can be better observed when we increase population size (Fig. 5). Bursts of rapid change are related to the evolutionary moments in which the pockets of isolated agents created by the late connectivity become connected. On the other hand, populations with mid, and, in particular, early connectivity dynamics followed a single-peak evolutionary dynamic. This is due to the fact that high-quality variants could spread in the system continuously (and without any additional restriction imposed by the connectivity dynamic) until the population reached its equilibrium.

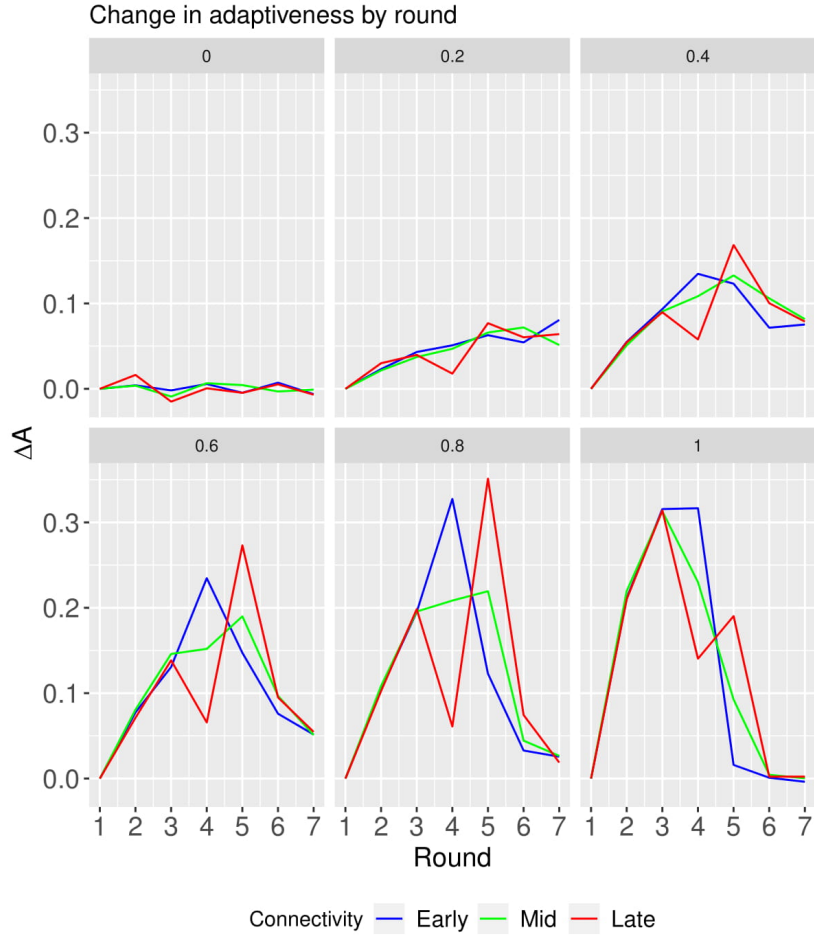

Figure 4. Change in adaptiveness $\Delta A$ of high quality variants by round, averaged over each level of connectivity and content bias in micro-societies of 8 agents. Results above 0 indicate that the proportion of high quality variants increased relative to the previous round. When $\Delta A = 0$, variant frequency was stable from round to round.

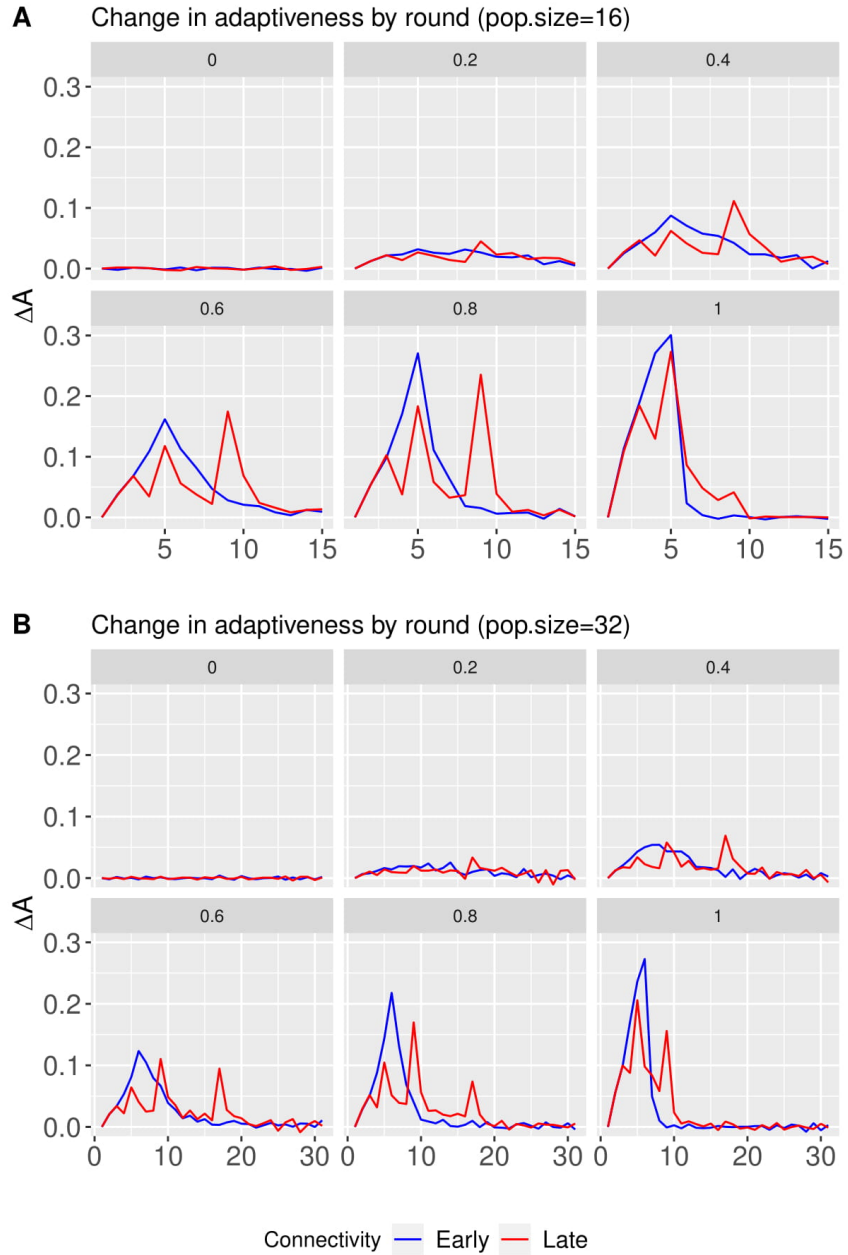

Figure 5. Change in adaptiveness $\Delta A$ of high quality variants by round, averaged over each level of connectivity and content bias, for population size=16 (A) and population size=32 (B). Results above 0 indicate that the proportion of high quality variants increased relative to the previous round. When $\Delta A = 0$, variant frequency was stable from round to round.

## 4. Discussion

Using a computational approach, the present study extends formal and experimental findings about learning in social networks by simulating pairwise interactive micro-societies where both individual cognitive biases, memory constraints and population connectivity dynamics are systematically manipulated. Our results show that connectivity dynamics affect the time-course of the spread of variants: when populations take longer to reach full connectivity, convergence onto a single cultural variant is slowed. We also show that content bias accelerates convergence and amplifies the effects of connectivity dynamics. Larger memory size and coordination bias, especially egocentric bias, are also shown to slow convergence. Finally, connectivity dynamics are shown to affect the frequency of high quality variants (adaptiveness), with late connectivity populations showing bursts of rapid change in adaptiveness followed by periods of relatively slower change, and early connectivity populations following a single-peak evolutionary dynamic.

Our computational results imply that the cognitive biases alone may not be sufficient to predict the rate of convergence of populations on shared cultural conventions: while content bias is a strong predictor of convergence, in some circumstances its effect can be modulated by the population connectivity dynamic (Figs. 2 and 3). This means that adding connectivity dynamics may improve the predictive power of models based on cognitive biases and social networks, especially in cases of strong content biases. Population convergence on shared cultural conventions is driven by agents' content bias, and the time required to reach a certain degree of convergence (or time to convergence) can be deeply affected by the specific order of interactions between agents, that is, by the population connectivity dynamic: in general, the less connectivity the more time is needed to converge. Furthermore, the effects of these different dynamics in the order of interactions of the agents can be observed even if we maintain the same network topology, in our case a fully connected network.

Our results also identify a general tendency for adaptiveness to change over time and for cultural variants to converge on high quality variants, in such a way that it is possible to identify causal links between connectivity dynamics and evolutionary trajectories. In this way, in late connectivity dynamics several punctuational bursts occur in the course of a complete cycle of interactions between agents. In contrast, early connectivity dynamics follow a single-peak evolutionary trajectory. These computational results extend a number of studies that, under a variety of assumptions, have proposed punctuational or rapid bursts of change as a feature of cultural and language evolution (Atkinson, Meade, Venditti, Greenhill & Pagel, 2008; Fitch, 2008; Janda & Joseph, 2003; Dixon & Robert Malcolm Ward, 1997; Sabherwal, Hirschheim, & Goles, 2001). Punctuational changes in our model may provide insight into processes underlying the human ability to adapt quickly to cultural variants introduced by new agents (e.g. due to migration), showing that these changes can be induced merely by manipulating the order of interactions in a population.

In high content-biased populations, the effect of the connectivity dynamic is amplified (Figs. 2 and 3), while coordination bias and memory size effects are masked (Figs. 6S, 7S and 8S). Interestingly, in low content-biased populations the effect of these parameters became visible: when compared with a drift model, egocentric and allocentric biases both reduce convergence (Fig. 3S, 5S). This is because cultural diversity can more easily be maintained over time in the presence of behaviors that maximize the occurrence of either self-produced signals (in the case of egocentric agents) or partner-produced signals (in the case of allocentric agents). The effect of egocentric bias is stronger than that of allocentric bias (Figs. 3S and 4S). This is due to the fact that fully egocentric agents stick to their own variant, which is always the same in the egocentric memory, unless there is an innovation. At the population level, this means that each agent has a different variant, returning maximum entropy (which can only decrease through mutation). Fully allocentric agents, in contrast, always adopt variants produced by others. High variation is thus maintained, but to a lesser extent than in the egocentric case because allocentric agents choose variants from among all the variants stored in their allocentric memories (variants produced by current or previous partners). This sometimes leads to more than two agents converging on the adoption of a variant, and therefore reducing entropy (Fig. 8S).

Memory also shows its effect more markedly when content bias is low or intermediate. The longer the memory span, the more variation is maintained, as variants from earlier rounds, that might not appear at one round, are kept in memory and may reappear. Our study is consistent with previous literature showing that memory limitations lead to a reduction in variation (Ferdinand et al, 2013; Tamariz & Kirby, 2015). Nevertheless, we show that this reduction could be masked in high content-biased populations, when agents have a strong preference for signals with high intrinsic value.

Our results also agree with recent studies showing that population structure and population interaction can be strong predictors of cultural evolution (Derex & Boyd, 2016; Creanza, Kolodny, & Feldman, 2017; Derex, Perreault & Boyd, 2018). In addition, our model shows that cognitive biases and population connectivity dynamics may interact in important ways. When content-biased populations evolve in high isolation (late connectivity dynamic), convergence is slower than when they evolve in moderate (mid connectivity dynamic) or low isolation (early connectivity dynamic). This is because more isolated subpopulations cannot benefit from wider cultural exchange during the first rounds, those in which agents are acquiring the basis of their culture and storing it into their respective memories. This suggests that population structure and, in particular, the connectivity dynamics of the population, can have important effects on cultural convergence and should be taken into account when it comes to research on the interactions between cognitive biases, network structures and cultural evolution.

## 5. Implications

In the cultural evolution of our societies, the structure of the population is important (Derex & Boyd, 2016; Björk & Magnusson, 2009; Hovers & Belfer-Cohen, 2006), and *inter-population connectivity* is more than a simple reflection of cultural accumulation (Creanza, Kolodny & Feldman, 2017). In a similar way, what our model shows on a micro scale, is that cultural isolation through population fragmentation interacts with individual-level biases, altering the way in which cultural evolution proceeds. In the light of evolutionary theory applied to cultural evolution (Atkinson, Meade, Venditti, Greenhill & Pagel, 2008; Fitch, 2008), our results also suggest that, in some scenarios, bursts of change in the cultural system may be partly explained by the order in which individuals interact over time. This demonstrates, for the first time, a direct connection between convergence, adaptiveness and population connectivity dynamics for a fixed range of combinations of individual cognitive biases.

In a population of interacting individuals, interactions between different actors may occur simultaneously and fragmentarily. For instance, a pattern may lead to temporary or permanent isolation of one or more sub-populations. Undoubtedly, in a comparable way, cultural mixing and fragmentation are determining factors in the formation of new cultural identities and languages during creolization (Thomason & Kaufman, 1992). According to our results, convergence processes in these situations might occur faster or slower depending on the degree of cultural contact and integration (large-scale effect of connectivity) whenever the subpopulations involved had shared a similar level of cultural preferences (which depends on the content bias).

As for everyday situations, there are many socio-cultural processes that are governed by turn-based interventions (Sacks, Schegloff & Jefferson, 1978), in which the organization of the turn taking might play an important role in, for example, the formation of sides during jury deliberations (Manzo, 1996), infant vocalizations (Bloom, Russell & Wassenberg, 1987), computer mediated communication (Garcia & Baker Jacobs, 1999) or communication in group decision-making (Bormann, 1996; Stasser & Taylor, 1991). In our study, we have shown that the interaction between connectivity dynamics and cognitive biases can affect turn-based cultural processes. Our computational model can be used to fit real data obtained from turn-based cultural processes and might be helpful to improve the organization of the turn taking by mitigating undesirable effects linked with one particular connectivity.

Our findings are consistent with scientific models, and theoretical and experimental studies of human communication showing that convergence is driven by content biases (Gong et al., 2008; Tamariz et al., 2014), and also agree with studies on rational learning in social networks showing that the level of convergence is partially determined by the degree of connectivity in the social network (Mueller-Frank, 2013, Olfati-Saber & Murray, 2004). However, we added an important socio-structural ingredient that interacts with the mechanism of alignment: the impact of the population connectivity dynamic. To our knowledge, this factor has not been taken into account in experimental work or models on cultural and language evolution, and it would be especially relevant to those researchers that use dynamic interactive microsocieties of agents switching partners over time (e.g. communicative games, cooperative games or tournaments: Fay et al. 2008, 2010; Tamariz et al. 2014; Mesoudi & Whiten, 2008; Caldwell & Smith, 2012; Baum, Richerson, Efferson & Paciotti, 2004; Byun, De Vos, Roberts & Levinson, 2018). In most cases, experimental designs of microsocieties of interacting actors only include one pair composition out of all the possible combinations of pair shuffling, and therefore, outcomes are related with only one specific population connectivity dynamic, potentially affecting the accuracy of the generalizations made by these studies. Our results suggest that this type of research would benefit from experimental designs that control the probability of occurrence of each possible connectivity dynamic.

Our agent-based model is a simplification of a specific problem. Each agent is characterized by a combination of biases towards the quality and origin of a set of variants. The network topology is complete and organized in dyads. Thus, following Ariel Rubinstein (2006), in the dilemma of responding to reality, we regard our model as a very limited set of assumptions which is inevitably inapplicable to many contexts. The utility of modeling in social sciences is to establish links between concepts and outcomes by creating an imaginary situation, not making extremely accurate predictions that are easily transferable to real-world settings. Thus, future research on the topic of this paper should go through experimentation with human participants, with the aim of verifying assumptions and conclusions. For example, further experiments with human participants could test the finding that different pairing setups can affect convergence and adaptiveness.

Properties of populations can be important predictors of cultural evolution, and our model has shown that convergence can be altered by the connectivity dynamic. This may help improve the experimental design of ongoing research in the field of cultural evolution and better explain the interactions between network topologies, cognitive biases and cultural transmission.

## Acknowledgments

We gratefully acknowledge the generosity of the Centre for Language Evolution (University of Edinburgh), without which the present study could not have been completed. We thank the Universitat Autònoma de Barcelona, Heriot-Watt University and the University of Western Australia for their support. We would also like to show our gratitude to Sergio Balari and Asha Sato for their comments during the course of this research.

## Data accessibility

Electronic supplementary material and simulation code are available at:
https://github.com/jsegoviamartin/network_connectivity_dynamics_model

## Authors' contributions

J.S.M. and M.T. designed the study. J.S.M. coded, performed and analysed simulations. J.S.M, M.T., N.F. and B.W. wrote the manuscript.

## Competing interests

The authors declare no competing interests.